\documentclass[aps,prd,groupedaddress,]{revtex4}
\usepackage{latexsym}
\usepackage[dvips]{graphicx}

\newcommand{\eq}{\begin{equation}}
\newcommand{\feq}{\end{equation}}
\newcommand{\eqn}{\begin{eqnarray}}
\newcommand{\feqn}{\end{eqnarray}}
\newcommand{\arr}{\begin{eqnarray*}}
\newcommand{\farr}{\end{eqnarray*}}
\newcommand{\beq}{\begin{equation}}
\newcommand{\eeq}{\end{equation}}
\newcommand{\bea}{\begin{eqnarray}}
\newcommand{\eea}{\end{eqnarray}}

\newcommand{\G}{{\cal G}}

\def\beq{\begin{equation}}
\def\eeq{\end{equation}}
\def\bea{\begin{eqnarray}}
\def\eea{\end{eqnarray}}
\def\bc{\begin{displaymath}}
\def\ec{\end{displaymath}}
\def\lb{\label}

\def\g{\gamma}
\def\G{\Gamma}
\def\de{\delta}
\def\ep{\varepsilon}

\def\ep{\epsilon}

\def\lb{\label}

\begin{document}

\preprint{INFNCA-TH0302}

\title{Entropy of the 3-brane and the  AdS/CFT duality from 
a two-dimensional perspective}

\author{Mariano Cadoni}
\email{mariano.cadoni@ca.infn.it}
\affiliation{Dipartimento di Fisica,
Universit\`a di Cagliari, and INFN sezione di Cagliari, Cittadella
Universitaria 09042 Monserrato, ITALY}


\begin{abstract}
A two-dimensional (2D) scalar-tensor gravity theory is used to describe the 
near-horizon near-extremal behavior of black 3-branes solutions 
of type IIB string theory. The asymptotic symmetry group of the 2D, asymptotically  
anti de Sitter (AdS) metric, 
is generated by a Virasoro algebra.
The (non-constant) configuration of  the 2D scalar field, which parametrizes  
the volume of the 3-brane, breaks the conformal symmetry and produces a 
divergent central charge in the 
Virasoro algebra. Using a renormalization  procedure we find a finite value
for the central charge and by means of the 
Cardy formula, the entropy of the black 3-brane in terms of 
microstates of the conformal field theory
living  on the boundary of the 2D spacetime. 
We find for the entropy as a function of the
temperature the power-law behavior $S\propto T^3$. 
Unfortunately, owing to a scale symmetry of the 2D model the 
proportionality constant is undetermined. 
The black 3-brane case is a nice example of how finite temperature effects
in higher dimensional AdS/CFT dualities can be described by
a AdS$_2$/CFT$_1$ duality endowed  with a scalar field
that breaks the conformal symmetry and produces a non-vanishing central charge. 
\end{abstract}


\maketitle

\section{Introduction}
The brane solutions of type II string theory in ten dimensions  
play a crucial role in the anti de Sitter/conformal field theory 
correspondence (AdS/CFT) \cite{Maldacena:1997re,Witten:1998qj,Aharony:1999ti}. 
Specializing to the   AdS$_{5}$/
CFT$_{4}$ case,  the low energy string theory splits into two 
decoupled pieces,  bulk supergravity and the near-horizon 
limit of the extremal 3-brane. In the near-horizon limit the 
extremal 3-brane has the  AdS$_{5}\times S^{5}$ geometry. On the 
other hand the same theory has an other low energy limit describing 
bulk supergravity and super Yang-Mills theory in four dimensions.
This fact led Maldacena to the conjecture that $U(N)$ super 
Yang-Mills in four dimensions is dual to type IIB string theory on 
AdS$_{5}\times S^{5}$ \cite{Maldacena:1997re,Aharony:1999ti}. 

Most of the progress about the string/gauge theory duality came from
comparison between the two theories at zero temperature, which for 
the supergravity side means  considering the extremal brane.
Finite temperature behavior, which is important both for testing the duality
and for discussing the thermodynamics of the brane from a 
microscopical point of view, can be discussed considering the 
near-extremal brane. 

Finite temperature breaks  conformal invariance. Moreover,  the 
discussion of  the strong  t'Hooft coupling regime  of the Yang-Mills
theory at finite temperature is problematic.
The entropy of the brane calculated using the usual 
Bekenstein-Hawking formula  differs from that calculated 
for the (weak-coupled) Yang-Mills theory by a proportionality factor
\cite{Gubser:1996de,Gubser:1998nz,Aharony:1999ti,Klebanov:2000me}.

The discrepancy is due to the fact that the gravity 
description is supposed to be valid at strong t'Hooft coupling.
The difficulty in performing calculations at strong t'Hooft coupling
makes it almost impossible to go beyond a qualitative explanation of 
the presence of the proportionality factor. In particular, it 
prevents a quantitative  explanation of the  Bekenstein-Hawking entropy 
of the brane in terms of microstates.
 
Two-dimensional gravity models, in particular dilaton gravity models, 
have been used with success in several contexts to discuss the 
near-horizon, near-extremal
limit of  higher dimensional black hole solutions of string theory
\cite{Callan:rs,Cadoni:1993rn,Cadoni:1999gh,Youm:1999xn,Grumiller:2002nm}.
They emerge in a natural 
way after dimensional reduction of 
the higher-dimensional gravity  theory and give a simple, effective 
description of the higher dimensional physics. It is therefore 
tempting to use them to describe the near-horizon, near-extremal limit 
of  black 3-branes. 
An other feature that makes  2D gravity models very interesting in this 
context is the relatively simple form that the AdS/CFT correspondence 
takes in two spacetime dimensions (AdS$_{2}$/CFT$_{1}$)
\cite{Cadoni:1999ja,Cadoni:1998sg,Caldarelli:2000xk,Cadoni:2000ah,
Cadoni:2000gm,Astorino:2002bj}. If the near 
extremal 
3-brane admits an effective near horizon description in terms of a 
2D gravity model, one can also hope that  the  (finite 
temperature) AdS$_{5}$/CFT$_{4}$ duality admits an effective 
description in terms of a AdS$_{2}$/CFT$_{1}$ duality.

This paper is devoted to the attempt of giving a two-dimensional 
description of both the near-horizon, near-extremal regime of black 3-branes 
of type II string theory and  of the  AdS$_{5}$/CFT$_{4}$  duality.
After a brief review of the black 3-brane solutions of type IIB 
string theory and the related AdS$_{5}$/CFT$_{4}$ duality (Sect. II ),
we proceed performing a dimensional reduction that produces a 2D 
dilaton gravity model describing the near-horizon, near-extremal 
behavior of the black 3-brane (Sect. III). We  study the group of 
asymptotical symmetries  of the 2D metric (which has asymptotical AdS 
behavior), show that it is generated by 
a Virasoro algebra and discuss its relationship with the conformal 
group in four dimensions (Sect. IV). The central charge of the Virasoro 
algebra diverges and we have to use a renormalization procedure to find
a finite value for it. Using the Cardy formula we calculate the entropy 
of the 3-brane by counting states in the boundary conformal field theory.
We reproduce the power-law behavior $S\propto T^{3}$ (Sect.  V).
Owing to a scale symmetry of the 2D model, which we discus in detail 
in Sect. VI and VII, the proportionality constant is undetermined.

\section{Black  3-branes and the AdS$_{5}$/CFT$_{4}$ duality }

Black 3-branes solutions of type IIB string theory in ten dimensions,
in particular their near horizon behavior, play a fundamental role in 
the formulation of the AdS$_{5}$/CFT$_{4}$ correspondence. In this section 
we will briefly review some known facts about these solutions.

Let us consider the low energy effective action for type IIB string 
theory in ten dimensions.  In the Einstein frame the action reads 

\beq\lb{saction}
S_{IIB}¥= {1\over (2 \pi)^{7} (\alpha')^{4} g_{s}^{2}}\int d^{10}x 
\sqrt{-g}\left(R - {1\over 2} (\nabla \Phi)^{2} - {2\over 
5!}F_{5}^{2}\right),
\feq
where $\alpha'$ is the string tension,  $g_{s}$ is the asymptotical 
string coupling constant, $\Phi$ is the dilaton and $F_{5}$ is the 
field strength for the 4-form potential, $F_{5}= dA_{4}$.
We are interested in solutions of the theory that are charged with 
respect to the Ramond-Ramond (RR) form $A_{4}$. They are given by  
(self dual) black 
3-branes with a constant dilaton, \cite{Horowitz:cd,Duff:1991pe,Duff:1994an} 
\beq\label{brane}
ds^{2}= H^{-1/2}\left(-f dt^{2}+ 
\sum_{i=1}^{3}dx^{i}dx^{i}\right)+H^{1/2}\left( f^{-1} 
dr^{2}+r^{2}d\Omega_{5}^{2}¥\right), \quad e^{-2\Phi}=g_{s}^{-2},
\feq
where
\beq
H(r)=1+{r_{-}^{4}\over r^{4}},\quad f(r)= 1-{r_{0}^{4}\over r^{4}}.
\feq
The solution describing a 3-dimensional electric source is 
characterized by the field strength
\beq\lb{fs}
F_{tijkr}= \epsilon_{ijk}H^{-2}{N\over r^{5}},
\feq
where $N$ is the RR charge. 
The dual magnetic configuration can be easily obtained from Eq. 
(\ref{fs}) applying the Hodge duality.
The parameters $r_{-}$ and $r_{0}$ are related to the mass per unit 
of volume and to the RR charge $N$ of the brane (see for instance 
\cite{Aharony:1999ti}).
The solution (\ref{brane}) describes a black brane with an event 
horizon at $r=r_{0}$. The brane becomes extremal for $r_{0}=0$.

We are particularly interested in the near-horizon 
behavior of the extremal brane solution (\ref{brane}). It is exactly this 
regime that is the relevant one for the AdS$_{5}$/CFT$_{4}$ correspondence.
In the near-horizon limit the extremal black brane has the geometry 
of AdS$_{5}$$\times$ S$^{5}$ \cite{Maldacena:1997re}. This fact led Maldacena 
to the conjecture that ${\cal{N}}=4$ $U(N)$ super Yang-mills theory in 
four dimensions is dual to type IIB superstring theory on  
AdS$_{5}$$\times $S$^{5}$. 
Most of the evidence about the AdS/CFT duality comes from comparison 
between quantities of the gauge and string theory that are protected by 
the conformal invariance (and by supersymmetry) of the  AdS$_{5}$ background. 
Excitations of the 3-brane above extremality in general 
break conformal invariance  and the brane acquires a finite 
temperature $T>0$. The near-extremal 3-brane  has a natural 
correspondence with the gauge field theory at finite temperature.   
The gravity solution describing the 3-brane in  the near-extremal,
near-horizon regime can be easily obtained from  Eq. (\ref{brane})
taking the $r\to 0$ limit, keeping both $r/\alpha'$ and the energy above 
extremality finite \cite{Maldacena:1997re,Aharony:1999ti}
\beq\label{extbrane}
ds^{2}= R_{0}^{2}\left\{ u^{2}\left[- \left(1- {u_{0}^{4}\over 
u^{4}}\right)dt^{2}+ 
\sum_{i=1}^{3}dx^{i}dx^{i}\right]+
{du^{2}\over u^{2}}\left(1- {u_{0}^{4}\over 
u^{4}}\right)^{-1}+d\Omega_{5}^{2}¥\right\}, 
\eeq
where $R_{0}¥^{4}¥=r_{-}^{4}=4\pi N 
(\alpha')^{2} g_{s}$, $u_{0}=r_{0}/R_{0}¥^{2}$ and the coordinate $u$ is 
related with the coordinate $r$ of Eq. (\ref{brane}) by $u=r/R_{0}¥^{2}$.
Moreover,  in the extremal limit  the RR field strength  is given by

\beq\lb{fs1}
F_{tijkr}= 2 R^{4}_{0}¥\epsilon_{ijk}u^{3}.
\feq
The temperature of the brane $T$  is given by 
\beq\lb{temp}
T= {u_{0}\over \pi}.
\feq
Working in the canonical ensemble, the Bekenstein-Hawking entropy 
$S$ and the energy of the excitation above extremality $E$ can be  
expressed as a function of the temperature \cite{Gubser:1996de},

\beq\lb{entropy}
S={\pi^{2}\over 2}V N^{2}T^{3},\quad E= {3\over 8}\pi^{2}V N^{2}T^{4},
\feq
where $V$ is the volume of the 3-brane.
In the spirit of the AdS/CFT correspondence the result 
(\ref{entropy}) should be matched by computations of the entropy in 
the finite-temperature $U(N)$ Yang-Mills theory. 
The gauge theory computation yields 
the result \cite{Gubser:1996de,Gubser:1998nz}
\beq\lb{entropy1}
S_{YM}={4\over 3}S_{brane}.
\feq
The origin of the factor $4/3$ is, qualitatively, well 
understood \cite{Klebanov:2000me}. The gauge theory computation is 
performed at zero 
t'Hooft coupling $Ng_{YM}^{2}$, whereas the gravity description given 
by the 3-brane is supposed to be valid at strong t'Hooft coupling.
A quantitative explanation of the result (\ref{entropy1}) is much more 
involved owing to the difficulty of doing calculations for the 
Yang-Mills theory at strong t'Hooft coupling.
In the next section we will try to address the problem using a simplified 
two-dimensional gravity model.

\section{Dimensional reduction}

The near-horizon, near-extremal solution (\ref{extbrane}) factorizes 
as a product of a, asymptotically AdS, five-dimensional (5D)  spacetime times 
a 5-sphere of 
radius $R_{0}¥$. We can obtain a effective 2D gravity model describing the 
near-horizon, near-extremal regime of the 3-brane by introducing a 
scalar field $\phi$, which parametrizes the volume of the 3-brane 
embedded in the 5D spacetime. We perform the dimensional reduction from ten 
to two dimensions using  for the metric the ansatz,
\beq\lb{dimred}
ds_{(10)}^{2}=ds_{(2)}^{2}+ \phi^{2/3}\sum_{i=1}^{3}dx^{i}dx^{i}
+ R_{0}¥^{2}¥d\Omega_{5}^{2}¥,\,
\feq
Notice that the volume of the 3-brane embedded in the 5D 
spacetime  is given by
\beq\lb{ev}
{\cal{V}}=\phi V.
\feq
The ansatz for the RR field strength and for the dilaton  follows 
directly from Eqs. (\ref{brane}), (\ref{fs1}),
\beq\lb{afs}
{F_{5}^{2}\over 5!}= \pm {4\over R_{0}¥^{2}},\quad  e^{-2\Phi}=g_{s}^{-2}.
\feq
where the $\pm$ sign refers to magnetic and electric solutions 
respectively. In the following we will consider only the dimensional 
reduction of the magnetically charged 3-brane.
The dimensional reduced action is obtained using Eqs. (\ref{dimred}) 
and (\ref{afs}) into the ten-dimensional action (\ref{saction}).
It has the form of a 2D dilaton gravity model,
\beq\lb{2daction}
S_{(2d)}¥= {\cal K}\int d^{2}x \sqrt{-g} \phi \left[ R +{2\over 3}
{(\nabla\phi)^{2}\over \phi^{2}}+ \lambda^{2}\right],
\feq
where ${\cal K}$, the inverse of the 2D Newton constant, is given by
\beq
{\cal K}={V N^{2}¥\over 8 \pi^{2}R_{0}^{3}},\quad   \lambda^{2}={12\over R_{0}^{2}}.
\feq

The 2D gravity model (\ref{2daction}) is a particular case of general 
class of models, which have been already investigated in the 
literature. The black hole solutions of the theory are given 
by \cite{Cadoni:2001ew}
\beq\lb{bh}
ds^{2}=- \left( b^{2}r^{2} - {A^{2}\over  b^{2}r^{2}}\right) dt^{2}+
\left( b^{2}r^{2} - {A^{2}\over  b^{2}r^{2}}\right)^{-1}¥ dr^{2},
\quad  \phi=\phi_{0}(br)^{3},
\feq
where $b=1/R_{0}$ and  $A, \phi_{0}$ are integration constants.

The 2D black hole solution gives an effective description of the 
near-extremal near-horizon regime of the 3-brane (\ref{extbrane}).
The parameters $A, \phi_{0}$  can be easily identified in terms of  
brane parameters. Changing in Eq. (\ref{bh}) the radial coordinate,
$r=uR_{0}^{2}¥$ and requiring the black hole solution (\ref{bh}) to match 
the 2D section of the black brane solution (\ref{extbrane}), 
one readily finds
\beq\lb{A}
A=u_{0}^{2}R_{0}^{2}=\left({r_{0}\over R_{0}}\right)^{2}.
\feq
The integration constant $\phi_{0}$ is related with a  on-shell
scale symmetry of the 2D gravity theory. Rescaling the scalar field 
$\phi$, $\phi\to\mu\phi$ the 2D action (\ref{2daction}) changes by an 
overall factor, $S_{(2d)}\to \mu S_{(2d)}$. We can use this classical 
symmetry of the action to change the normalization of the action 
(\ref{2daction}) to ${\cal K}=1/2$ (the normalization used in Ref.
\cite{Cadoni:2001ew}). This fixes the value of $\phi_{0}$,
\beq\lb{phi0}
\phi_{0}={N^{2}V\over 4\pi^{2}R_{0}¥^{3}}.
\feq
Let us now compute the thermodynamical parameters associated with the 
2D black hole and show that they reproduce exactly those of the 3 black 
brane. The ADM mass $M_{bh}$, temperature $T_{bh}$ and entropy 
$S_{bh}$ are \cite{Cadoni:2001ew},
\beq\lb{tp}
M_{bh}= {3\over 2}\phi_{0}b A^{2},\quad T_{bh}={b\over \pi} 
\sqrt{A},\quad S_{bh}=2 \pi \phi_{0}A^{3/2}.
\feq
Using Eqs (\ref{A}),(\ref{phi0}) and comparing the previous equations 
with Eqs. (\ref{temp}), (\ref{entropy}) one easily finds 
$T_{bh}=T_{brane}$,\, $S_{bh}=S_{brane}$ and $M_{bh}=E_{brane}$,
where $E_{brane}$ is  the energy of the brane above 
extremality. 
This result can be easily understood. The thermodynamics of 
both the 2D black hole and of the black 3-brane is determined by  the
behavior at the horizon, $r=(\sqrt{A})/b$. On the other hand this behavior is 
determined by the 2D sections of the metric, which are the same 
for the 2D black hole and the 3-brane.
The story looks different if we want to give a microscopical 
interpretation of the thermodynamical behavior or if we want to 
consider the region where the dual gauge theory is weak coupled. 
In both cases we need to move away from the horizon and to consider 
the asymptotical region $r\to\infty$. The natural framework for doing 
this is to investigate the asymptotical symmetries of the 2D solution.
This will be the subject of the next section.

\section{Asymptotical symmetries}
The group of asymptotical symmetries of the metric (\ref{bh}) have 
been already investigated in Ref. \cite{Cadoni:2001ew}. Consistently with 
the fact 
that the metric is asymptotically AdS, it was found  that the 
asymptotic symmetries  are 
generated by operators $L_{m}$ spanning a Virasoro algebra. 
Unfortunately, the central charge of the Virasoro algebra turns out 
to be divergent.  This divergence is essentially due to the 
asymptotical behavior of the scalar field $\phi\sim r^{3}$ (\ref{bh}). 
In  this paper we will use an alternative approach. We will 
renormalize the central charge by subtracting the divergent 
contribution of the background.

Changing the radial coordinate, 
\beq\lb{cc}
b^{2}r^{2}¥\to b^{2}r^{2}+ A,
\feq
the solution (\ref{bh}) becomes
\beq\lb{bh1}
ds^{2}=- \left( b^{2}r^{2} + 2A\right) \left(1+ {A\over  b^{2}r^{2}}
\right)^{-1}dt^{2}+
\left( b^{2}r^{2} +2A \right)^{-1}¥ dr^{2},
\quad\quad \phi=\phi_{0}\left[(br)^{2}+A\right]^{3/2}.
\feq
Expanding the solution near $r\to\infty$ we get,
\eqn\lb{aexp}
g_{tt}&=& -b^{2}r^{2}-A +{A^{2}\over b^{2}r^{2}} +{\cal O}(r^{-4})\,,
\nonumber\\
g_{rr}&=& {1\over b^{2}r^{2}}-{2 A\over b^{4}r^{4}}+{4 A^{2}\over b^{6}
r^{6}} +{\cal O}(r^{-8}),\,\\
\phi&=&\phi_{0}\left(b^{3}r^{3} +{3\over 2}Abr  +{3\over 8}{A^{2}\over
br} +{\cal O}(r^{-3})\right).\,\nonumber
\feqn
The metric is asymptotically AdS.  We are  led to impose the following
$r\to \infty$ boundary conditions 
\eqn\lb{bc}
g_{tt}&=& -b^{2}r^{2}+\g_{tt}¥ +{\G_{tt}¥\over b^{2}r^{2}} +{\cal O}(r^{-4})\,,
\nonumber\\
g_{rr}&=& {1\over b^{2}r^{2}}+{\g_{rr}¥\over b^{4}r^{4}}+{\G_{rr}¥\over b^{6}
r^{6}} +{\cal O}(r^{-8}),\,\nonumber\\
g_{tr}&=&{\g_{tr}¥\over b^{3}r^{3}}+{\cal O}(r^{-5}),\,\\
\phi&=&\phi_{0}\left(\rho (br)^{3} +\g_{\phi\phi}¥br  +
{\G_{\phi\phi}\over br}¥+{\cal O}(r^{-3})\right)\,\nonumber
\feqn
Where $\g,\G,\rho$ are deformations of the metric and of the scalar field.
The asymptotic symmetry group (ASG) of the metric is given by 
2D diffeomorphisms  which preserve the boundary conditions (\ref{bc}).
The associated Killing vectors are
\beq\lb{kill}
\chi^{t}=\ep(t) + {\ddot \ep(t)\over 2b^{4}r^{2}}+{\cal 
O}(r^{-5}),\quad
\chi^{r}=- \dot\ep(t)r + {\cal O}(r^{-2}),
\feq
where $\ep(t)$ is an arbitrary function of the time $t$. 
The leading terms of the Killing vectors (\ref{kill}) are typical
of every 2D metric with AdS asymptotic behavior 
\cite{Cadoni:1998sg,Cadoni:2001ew}. 

The generators $L_{k}$ of the 
 ASG satisfy the Virasoro algebra,
\beq\lb{vira}
[L_{k},L_{l}]=(k-l)L_{k+l}+{c\over 12}(k^{3}-k) \delta_{k+l,0},
\feq
where we allow for a nonvanishing central charge $c$.
The ASG has a natural realization in terms of the conformal group in 
one dimension (the diff$_{1}$ group) leading in a natural way to a 
AdS$_{2}$/CFT$_{1}$ duality \cite{Cadoni:1999ja}.
Notice that whereas the leading terms of the 2D metric are invariant
under the transformations generated by the Killing 
vectors (\ref{kill}), that of the scalar field are not.
The non-constant configuration of the scalar field $\phi$ breaks the 
the conformal symmetry of the metric. This breaking is a general 
feature of 2D AdS solutions endowed with a non-constant scalar field 
and it is the source of a non-vanishing central charge in the 
Virasoro algebra \cite{Cadoni:2000ah}.

Using the boundary conditions (\ref{bc}) and the Killing vectors 
(\ref{kill}) we  can easily find the transformation laws of the boundary 
fields $\g,\G,\rho$ under the action of the conformal group.
In the following we will make use only of the transformation laws for
$\g_{rr},\g_{tr}¥,\rho,\g_{\phi\phi}$. They are:
\eqn\lb{tl}
\delta \g_{rr}&=& \ep \dot \g_{rr}¥ +2 \dot\ep\g_{rr},\qquad\quad
\delta \g_{tr}= \ep \dot \g_{tr}¥ +3 \dot\ep\g_{tr}-{\ddot\ep\over b}
(\g_{rr}+\g_{tt}),\,\nonumber\\
\delta \rho&=& \ep \dot \rho -3 \dot\ep \rho,\qquad\quad
\delta \g_{\phi\phi}= \ep \dot \g_{\phi\phi}-\dot\ep\g_{\phi\phi}
+ {\ddot\ep\dot\rho\over 2b^{2}}.
\feqn

Let us now discuss the relationship between the ASG of the 2D metric 
(\ref{bh1}) and the isometry group of the extremal 3-brane (AdS$_{5}$).
The isometry group of  AdS$_{5}$ is $SO(2,4)$, which 
is locally isomorphic to the conformal group in four dimensions.
The dimensional reduction of the previous section applied to 
AdS$_{5}$ produces the 2D background solution (\ref{bh}) with $A=0$,
\beq\lb{bg}
ds_{b}¥^{2}= -(br)^{2}dt^{2}+(br)^{-2}dr^{2},\quad \phi_{b}=\phi_{0}(br)^{3},
\feq
describing AdS$_{2}$ endowed with a non-constant field $\phi$.
Owing to the scalar character of $\phi$, the configuration (\ref{bg}) 
is not invariant under the full conformal group $SO(2,4)$ but only 
under its subgroup generated by time-translations.
The dimensional reduction breaks the full isometry group of 
AdS$_{5}$. However, if we forget the field $\phi$, and concentrate 
ourselves on the metric part of the solution (\ref{bg}) we see that the
isometry group of the AdS$_{5}$ metric has been promoted by the 
dimensional reduction to the diff$_{1}$ group generated by the 
Killing vectors (\ref{kill}). Of course, this symmetry is broken by 
the non-constant value of $\phi$, but as we shall see in the next 
section the effect of this breaking is the appearance of a non-vanishing 
central charge in the Virasoro algebra (\ref{vira}).
We see here at work a very nice mechanism: the dimensional reduction 
allows us to describe the AdS$_{5}$/CFT$_{4}$ duality in terms of 
an AdS$_{2}$/CFT$_{1}$ duality in which the conformal symmetry is 
broken by a non-constant value of a scalar field.

\section{Central charge and entropy}

The central charge appearing in the Virasoro algebra (\ref{vira}) can 
be calculated using a canonical realization of the ASG 
\cite{Cadoni:1998sg,Cadoni:2001ew}.
Using the parametrization of the metric,

\beq
ds^{2}=-N^{2}dt^{2}+\sigma^{2}\left(dr+N^{r}dt\right)^{2}¥,
\feq
one can easily derive the Hamiltonian $H$ of the theory \cite{Cadoni:2001ew}.  
Boundary 
term $J$ (the charges) must be added to $H$ in order to have well-defined 
variational derivatives . On the $r\to\infty$ boundary of 
2D AdS spacetime the variation $\delta J$ is given by
\cite{Cadoni:2001ew}
\beq\lb{var}
\delta J=-\lim_{r\to\infty}\left[N\left(\sigma^{-1}\delta \phi' -
\sigma^{-2} \phi'\delta \sigma -{2\over 3} \sigma^{-1}\phi^{-1}\phi'
\delta \phi\right)-N'\sigma^{-1}\delta \phi +N^{r}\left( 
\Pi_{\phi}\delta\phi-\sigma\delta\Pi_{\sigma}\right)\right],
\feq
where $\Pi_{\sigma}$ and  $\Pi_{\phi}$ are momenta conjugate to 
$\sigma$ and $\phi$ respectively and the prime denotes derivative with 
respect to $r$. The central charge in the Virasoro algebra (\ref{vira}) 
can be calculated introducing time-integrated charges \cite{Cadoni:1998sg}
\beq\lb{intc}
\hat J={b\over 2\pi}\int_{0}^{2\pi/b}J dt,
\feq 
and using the commutator 
\beq\lb{alg}
\hat{\delta_{\omega}J(\epsilon)}=\left[\hat J(\epsilon), \hat J(\omega)\right].
\feq

If we follow Ref \cite{Cadoni:2001ew} and  compute the central charge using 
Eq. (\ref{var}) we get a divergent result. The divergent contribution 
is due to the cubic divergent term in the scalar field $\phi$ 
(\ref{aexp}). More physically, it is due to the fact that arbitrary 
excitations 
of the $r=\infty$ boundary have infinite energy \cite{Cadoni:2001ew}. In Ref 
\cite{Cadoni:2001ew} we did not try to subtract the divergent part of the charges 
in order to get a renormalized, finite answer.  
Here we will use an alternative approach. We will show that the 
charges $J$ can be consistently renormalized by subtracting the 
divergent contribution of the background.

Let us first compute the energy $M$ of the excitations on the $r\to\infty$ 
boundary. This can be done either evaluating $J(\epsilon=1)$ from Eq.
(\ref{var}) or using Mann's formula for the mass \cite{Mann:1992yv}. In both 
cases we find
\eqn\lb{mann}
M&=&2 b\phi_{0}\rho^{4/3}\left\{\left[{1\over 12 b^{2}}¥\left({\dot\rho\over 
\rho}\right)^{2}¥+
{\g_{\phi\phi}\over \rho}+ {3\over 4}\g_{rr}\right](br)^{2}
+ \left[2 {\G_{\phi\phi}\over \rho} +
{3\over 4}\left(\G_{rr}-\g_{rr}^{2}\right) +\right.\right.\nonumber\\ 
&+&\left.\left.{1\over 6} {\dot\g_{\phi\phi}\dot\rho\over \rho^{2}¥} -
{1\over 4}{\g_{tr}\dot\rho\over \rho} -{1\over 18}
{\g_{\phi\phi}\dot\rho^{2}¥\over \rho^{3}}
-{1\over 12}{\g_{tt}\dot\rho^{2}¥\over \rho^{2}}\right]\right\}
+{\cal O}(r^{-2}),
\feqn
For arbitrary deformations $\g,\G,\rho$ $M$ diverges quadratically 
as $r\to\infty$. However,  the quadratically divergent term cancels 
if we consider only on-shell deformations. In fact evaluating the 
field equations coming from the 2D action (\ref{2daction}) on the 
boundary conditions (\ref{bc}) one finds that $\g,\rho$  must 
satisfy the equation \cite{Cadoni:2001ew}
\beq\lb{const}
{1\over 12 b^{2}}\left({\dot\rho\over \rho}\right)^{2}¥+
{\g_{\phi\phi}\over \rho}+ {3\over 4}\g_{rr}=0.
\feq

The previous observations indicate the way to keep the energy of the 
boundary excitations  finite: one just needs to consider on-shell 
deformations. Unfortunately, this is not enough to guarantee the 
finiteness  of the other charges and of the central charge. 
The cubic asymptotic 
behavior of the scalar field $\phi$ is also responsible for the 
divergence of the other charges. Owing to its scalar character 
a non-constant configuration of $\phi$ always breaks the asymptotic 
symmetry of the metric. This is a general feature of 2D 
scalar-tensor theories of gravity, which plays an important role for 
the determination of the central charge in the Virasoro algebra. In 
these models  the origin of the central charge 
can be traced back to the breaking of the conformal symmetry due to a
non-constant value of $\phi$ \cite{Cadoni:2000ah}.  As long as the field $\phi$
depends linearly on $r$ the central charge remains finite \cite{Cadoni:2001ew}.
A behavior $\phi\sim r^{h}$ with $h<1$ implies a vanishing central 
charge, whereas for $h>1$ the central charge diverges.

The divergence of the central charge for $h=3$ can be easily cured.
The asymptotic AdS background configuration (\ref{bg})
is, strictly speaking, not invariant under the action of the ASG 
generated by the Killling vectors (\ref{kill}), owing to the 
non-constant value of $\phi$. It is therefore natural to subtract to 
the charges $J$ defined by Eq (\ref{var}) the contribution $J_{b}$ 
obtained evaluating $J$ on the background (\ref{bg}).
We are therefore led to define renormalized charges $J_{R}$,
\beq\lb{rc}
J_{R}=J-J_{b}.
\feq

Using Eqs (\ref{var}),(\ref{rc}) and computing the variations near the 
classical solution (\ref{aexp}),  we find after some manipulations
\beq\lb{e1}
\delta J_{R}(\epsilon)= \epsilon \delta M + \delta J_{1},
\feq
where $M$ is given by Eq. (\ref{mann}) and 
\beq\lb{e2}
\delta J_{1}={\phi_{0}\over b}\left(\dot \ep\de\dot \g_{\phi\phi}+
{3\over 4}\ddot \ep\de \g_{rr}-3A \ddot \ep\de\rho- 
{3\over 2} A\dot \ep\de\dot\rho-3 b \dot \ep \de \g_{tr}\right).
\feq
Considering only on-shell deformations $M$ and $J_{R}$ are
finite. The  central 
charge of the Virasoro algebra is determined  by the term  $\de J_{1}$,
through equation (\ref{alg}).

Evaluating  
Eq. (\ref{e2}) on the classical configuration (\ref{aexp}),
using Eq. (\ref {tl}),
and taking into account  that the charges  
$\hat J$ (\ref{intc}) are defined only up to a total time derivative,
 we find
\beq\lb{e4}
\de_{\omega}J_{1}(\ep)= {6\phi_{0}A\over b}\left(\dot\omega\ddot\ep- 
\dot\ep\ddot\omega\right).
\feq
Expanding in Fourier modes and using Eq. (\ref{alg}) 
we find the central charge $c$ of the 
Virasoro algebra
\beq\lb{central}
c=144 \phi_{0}A.
\feq
The central charge of the Virasoro algebra depends on the mass of the 
2D black hole. In fact from Eq. (\ref{tp}) we find 
$c\propto \sqrt{M_{bh}}$. 

Using the Cardy formula \cite{Cardy:ie},
\beq\lb{cardy}
S= 2\pi \sqrt{c l_{0}\over 6},
\feq
we can calculate the statistical entropy  $S$ associated with the 
boundary conformal theory. The eigenvalue $l_{0}$ of the Virasoro 
operator $L_{0}$ can be given in terms of the black hole mass $l_{0}=
M_{bh}/b=(3/2) A^{2}\phi_{0}$. The statistical entropy turns out to be
\beq\lb{sentropy}
S= 12 \pi \phi_{0} A^{3/2}.
\feq
We can now express, by means of Eqs. (\ref{phi0}), (\ref{tp}) 
the entropy either as a function of the temperature 
(canonical ensemble) or as function of the  energy 
(microcanonical ensemble)
\beq\lb{e5}
S= 3 \pi^{2} N^{2} V T^{3},\quad S= 4\, (6)^{1/4} \sqrt{\pi} \left(N^{2} V 
\over R^{3}_{0}¥\right)^{1/4}(ER_{0}¥)^{3/4}.
\feq
The statistical entropy associated with the conformal field theory 
living in the boundary of the 2D spacetime reproduces correctly the 
power-law behavior $S\propto T^{3}$ (or $S \propto E^{3/4}$ for the 
microcanonical ensemble) of the 3-brane (\ref{entropy}) and of the $U(N)$ 
gauge theory (\ref{entropy1}). This is a highly nontrivial, and to some extend 
unexpected, result. In previous calculations for a different kind of 2D 
AdS spacetime we have found the scaling behavior $S\propto T$, which is 
typical for a 2D CFT \cite{Cadoni:1998sg}. Our result indicates that the conformal 
field theory living on the boundary of 2D AdS 
spacetimes endowed with a  scalar field behaving asymptotically 
as $\phi\sim r^{d-1}$ can reproduce the 
entropy/temperature thermodynamical relation of  conformal field 
theories in $d\ge 2$ dimensions. The explanation of this peculiar feature
can be found in the breaking of the conformal symmetry  (the ASG) of 
the 2D AdS spacetime caused by the nonconstant scalar field $\phi$.

The proportionality factor between $S$ and $T^{3}$ in Eq. (\ref{e5})
deserves also some surprise. It is not the same as that associated with 
the  horizon of the   black 3-brane (or 2D black hole) (\ref{entropy}).
This was somehow to be expected because the theory on the horizon is 
strongly coupled whereas that on the $r\to \infty $ region is weakly 
coupled.  For this reason the most natural result for the entropy 
should have been  Eq. (\ref{entropy1}), i.e the entropy of the 
Yang-Mills theory at zero t'Hooft coupling.  This is not the case 
because of a factor 6. In the next section we will argue that the 
2D theory can only determine the scaling behavior $S\propto  T^{3}$
whereas the proportionality factor is from the 2D point of view 
essentially undetermined.

\section{Scale symmetry and  renormalization}
We have already noted in Sect. III that the 2D action (\ref{2daction}) has a 
classical symmetry $\phi\to \mu\phi$ that rescales the 2D Newton 
constant. More in detail, this scale symmetry appears as a 
subgroup of the isometry group of AdS$_{2}$. It is easy to realize 
that the scale transformation
\beq\lb{st}
r\to \mu r,\quad t\to \mu^{-1}t¥,
\feq
leaves the AdS metric (\ref{bg}) invariant, whereas the scalar $\phi$ 
scales as a three-dimensional volume
\beq\lb{ds}
\phi\to\mu^{3}\phi.
\feq
Hence the on-shell symmetry that rescales the 2D Newton constant can 
be identified as the scale isometry of AdS$_{2}$¥ Again, it is the non 
constant value of the scalar $\phi$ which is responsible for 
the breaking of  the isometry group of AdS$_{2}$, whose effect is a 
change of the 2D Newton constant.
As a consequence of this symmetry  $\phi_{0}$ 
is a sort of sliding field, which at the classical level remains 
undetermined. 
One could argue that this feature does not affect 
computation of the ratio
$S^{(a)}/S^{(h)}$ between  the entropy associated  with the horizon 
and the entropy of the CFT living in the asymptotic region, because 
one can use for calculating both $S^{(a)}, S^{(h)}$ the same classical solution.
This is not the case because in order to have a finite  central charge
we have to subtract the divergences. The renormalization procedure of 
the previous section is equivalent to define a renormalized  scalar
field
$\phi_{R}=\phi -\phi_{b}$.
This introduces the usual ambiguities related with the 
renormalization scheme. In general we are not allowed anymore to use 
the same $\phi_{0}$ for computations at the horizon and computations
in the asymptotic region.

For the above reasons the numerical factor appearing in Eq.
(\ref{central}) has no physical meaning, it can be changed simply by using 
an isometric transformation of AdS$_{2}$. Consequently, all the 
thermodynamical relations derived at the end of the previous section
are determine up to a unknown dimensionless proportionality factor.
This behavior is very similar to what has been observed in 2D  cosmological 
context \cite{Cadoni:2002rr}. Investigating the relationship between the Cardy 
formula and the Friedmann equation for 2D de Sitter gravity, we found 
a proportionality 
relation between the central charge and the inverse 
of the  Newton constant. Owing to the scale symmetry of the theory 
the proportionality factor is undetermined.

Although classically undetermined, $\phi_{0}$, or better the ratio
$\phi_{0}^{(a)}/\phi_{0}^{(h)}$ between its asymptotic value and that at
the horizon, could be determined at the quantum level. For instance 
one could use the sigma-model formulation  of  2D dilaton gravity 
\cite{Cavaglia:1998xj}, to compute the running of $\phi_{0}$. As observed in 
Ref. \cite{Cadoni:2001ew}, this is far from being trivial. In the $r\to\infty$ 
asymptotic region the sigma model is weakly coupled and one can use 
a perturbative approach, but near the horizon the theory 
becomes strongly coupled so that usual perturbative calculations become 
useless. 

Until now our discussion was based on arguments coming from the 2D 
gravity theory. These arguments have a nice formulation in terms of 
the 3-brane 
and of the AdS$_{5}$/CFT$_{4}$ correspondence.
The scale transformation (\ref{st}) appears also as isometry of 
AdS$_{5}$ ( solution (\ref{extbrane}) with $u_{0}=0$). In the 5D 
context the scale transformation  (\ref{st}) is an exact symmetry of 
the background solution if the coordinates $x^{i}$ of the 3-brane 
transform as $ x^{i}\to \mu^{-1}x^{i}$. This is the so-called 
ultraviolet/infrared (UV/IR) connection  \cite{Susskind:1998dq}. 
Points near the $r=0$ origin of the AdS spacetime  are mapped into 
the infrared region of the Yang mills theory.
In the 2D theory the brane is compactified and its volume is described 
by the scalar field $\phi$, which encodes the information 
about the embedding of the 3-brane in the 5D spacetime.
The scale symmetry is broken and the 
change of the scalar field $\phi$ under  scale transformations
can be describe as a change of the parameter $\phi_{0}$, 
$\phi_{0}\to \mu^{3}\phi_{0}$. Thus, running from the UV to the IR 
in the  Yang mills theory is equivalent to  running of the  2D 
Newton constant. In particular, because $\phi_{0}\propto G_{N}^{-1}$,
the UV region of Yang Mills theory correspond to the weak-coupled 
regime of the 2D gravity theory, whereas the IR region is in 
correspondence with the strong-coupled regime.

There is evidence that also the near-horizon region of the 2D black hole 
can be described by a CFT \cite{Carlip:2002be}. Therefore the 3-brane geometry  is 
characterized by two CFT's. One is generated by the ASG of the 
AdS$_{2}$ metric and the other is associated with the black hole horizon.
If this is true the running of the 2D Newton constant can be interpreted 
as the running of the central charge when moving from a conformal 
point to the other. 
This 2D feature has a counterpart in the large $N$ behavior of 
conformal field theories, for which relevant double-trace deformations produce a 
flow from an ultraviolet to an infrared  fixed point \cite{Gubser:2002vv}.

\section{Scaling of the thermodynamical parameters}
In the previous section we have considered the realization of the 
scale symmetry for the vacuum AdS$_{2}$ solution (\ref{bg}). The 
vacuum solution is characterized by zero mass, temperature and entropy. 
Physically, the scale symmetry means that we can change the size  of 
the system  without changing the thermodynamical parameters.
This is not true anymore when we consider the solution at finite 
temperature, i. e. the 2D black hole (\ref{bh}) (corresponding in the 
ten-dimensional theory to the black brane (\ref{extbrane}). We expect now the 
scale symmetry to be broken and that changing the size of the system 
will affect $M,T,S$. Nonetheless, we will see that the scale symmetry 
determines the scaling behavior of $M,T,S$. This is very similar to 
what has been discussed for a class of 2D dilaton gravity model in Ref.
\cite{Cadoni:2001tb}. 

The black hole metric (\ref{bh}) is not invariant under the scale transformation
(\ref{st}). However it can be made invariant if we change in Eq. 
(\ref{bh}) together 
with $r,t$ also the parameter $A$, $A\to\mu^{2} A$. 
From Eqs. (\ref{tp}) it follows immediately the scaling of the 
thermodynamical parameters
\beq\lb{stp}
T\to\mu T,\quad M\to\mu^{4} M, \quad S\to \mu^{3} S.
\feq
The scale transformation for  $\phi$, $\phi\to \mu^{3}\phi$ gives the 
scaling law for the volume $\cal{V}$ of Eq. (\ref{ev}),
${\cal{V}}\to \mu^{3} {\cal{V}}$, which is the expected one for  the volume 
of a three-dimensional object. 
Also the entropy $S$ scales extensively as a volume, whereas the mass 
$M$ is non-extensive.  

The previous scaling laws contain all the information necessary to 
determine up to a proportionality factor the thermodynamical 
behavior of the system. Writing ${\cal{V}}= \mu^{3}{\cal{V}}'$, we get from 
Eq. (\ref{stp}) $M({\cal{V}})= \mu^{4}M({\cal{V}}'),\,\
S({\cal{V}})= \mu^{3}S({\cal{V}}') $. Setting 
${\cal{V}}'=1$, trading ${\cal{V}}$ for $S$ and using the equation 
$dM=TdS$ we get the mass/entropy and temperature/entropy relations 
\beq\lb{rel}
M=\alpha S^{4/3},\quad T=\alpha {4\over 3}S^{1/3},
\feq
where $\alpha$ is a proportionality constant, which cannot be determined
using only scaling arguments. Again, 2D arguments   enable us to 
determine the exponents of the power-law behavior but not the 
proportionality constant.
Owing to the indetermination of $\alpha$,  
Eq. (\ref{rel}) hold both for the near horizon 3-brane solution (se  
Eqs (\ref{temp}), (\ref{entropy})) and for the Yang Mills Theory 
(see Eq. (\ref{entropy1}).
We can also eliminate $\alpha$ from the two Eqs (\ref{rel})
\beq\lb{nonext}
M={3\over 4}TS.
\feq
From the 2D point of view this equation can be understood as a 
violation of the Euler identity $M=TS$. It is a consequence of the non
extensivity of the  thermodynamical system. 

\section{Conclusion}

In this paper we have shown that one can use an effective 
two-dimensional description for the near-extremal near-horizon behavior
of the black 3-brane of type IIB string theory. This effective 
description allows us to formulate the AdS$_{5}$/CFT$_{4}$ duality at 
finite temperature as 
a AdS$_{2}$/CFT$_{1}$ duality with the conformal symmetry broken by 
nonconstant configuration of the scalar field parametrizing the volume
of the 3-brane. The breaking of the symmetry produces a  nonvanishing 
central charge. In particular, this allows for a computation of the 
entropy of the 3-brane in terms of the conformal field theory living 
in the boundary of AdS$_{2}$.
Although we could reproduce the 
entropy/temperature power-law behavior of the 3-brane a dimensionless 
proportionality constant, related with the 2D Newton constant,
remains undetermined.
This indetermination seems a general feature of 2D gravity solutions 
with AdS (or de Sitter) behavior \cite{Cadoni:2002rr,Cadoni:2001tb} 
and is due to  a scale symmetry of the 
model. From the 3-brane point of view  this scale symmetry is 
related with the UV/IR connection in the AdS$_{5}$/CFT$_{4}$ 
duality.

The presence of an undetermined dimensionless constant 
results in a loss of predictive power of the two-dimensional model.
In particular, it makes impossible, at least at the level of our present 
investigation, the determination of the ratio (\ref{entropy1}) between 
the Bekenstein-Hawking entropy of the black 3-brane and the entropy of 
the Yang-Mills theory. Qualitatively this ratio can be explained in 
the 2D gravity theory in terms of the running of the Newton constant 
(or equivalently of the central charge) when moving from one to the 
other conformal point of the geometry.  
It is very difficult to go beyond  this qualitative analysis. The  
theory is  weakly coupled  at the one
conformal point ( the $r\to\infty$ asymptotical region) 
but becomes 
strongly coupled  at  the other (the horizon).

\begin{acknowledgments}
We are very grateful to P. Carta and S. Mignemi for interesting discussions 
and useful comments.
\end{acknowledgments}

\end{document}